\documentclass[preprint]{aastex63}

\usepackage[utf8]{inputenc}
\usepackage{amsmath, color, amssymb, latexsym, graphicx, lipsum}
\newcommand{\ua}{{\bf u}_{\text{A}}}
\newcommand{\uh}{{\bf u}}
\def\ADD#1{{\textcolor{black}{#1}}}   % addition for comments!

\submitjournal{ApJ}

\shorttitle{Solar Wind Turbulence Around Mars}
\shortauthors{Andr\'es et al.}

\begin{document}

\title{Solar Wind Turbulence Around Mars: Relation Between The Energy Cascade Rate And The Proton Cyclotron Waves Activity}

\correspondingauthor{Nahuel Andr\'es}
\email{nandres@iafe.uba.ar, nandres@df.uba.ar}

\author[0000-0002-1272-2778]{Nahuel Andr\'es}

\affiliation{Instituto de Astronom\'ia y F\'{\i}sica del Espacio, CONICET-UBA, Ciudad Universitaria, 1428, Buenos Aires, Argentina}
\affiliation{Departamento de F\'{\i}sica, UBA, Ciudad Universitaria, 1428, Buenos Aires, Argentina}

\author[0000-0001-9210-0284]{Norberto Romanelli}
\affiliation{Solar System Exploration Division, NASA Goddard Space Flight Center, Greenbelt, MD, USA}
\affiliation{CRESST II, University of Maryland, Baltimore County, Baltimore, MD, USA.}

\author[0000-0002-8587-0202]{Lina Z. Hadid}
\affiliation{LPP, CNRS, École polytechnique, Institut Polytechnique de Paris, Sorbonne Université, F-91128 Palaiseau, France}
\affiliation{European Space Agency, ESTEC, Noordwijk, Netherlands}

\author{Fouad Sahraoui}
\affiliation{LPP, CNRS, École polytechnique, Institut Polytechnique de Paris, Sorbonne Université, F-91128 Palaiseau, France}

\author[0000-0002-2778-4998]{Gina DiBraccio}
\affiliation{Solar System Exploration Division, NASA Goddard Space Flight Center, Greenbelt, MD, USA}

\author[0000-0001-5258-6128]{Jasper Halekas}
\affiliation{Department of Physics and Astronomy, University of Iowa, Iowa City, Iowa, USA}

\begin{abstract}
The first estimation of the incompressible energy cascade rate at magnetohydrodynamic (MHD) scales in the plasma upstream of the Martian bow shock is obtained, making use of magnetic field and plasma observations provided by Mars Atmosphere and Volatile EvolutioN (MAVEN) over 600 orbits. In particular, the energy cascade rate is computed for events with and without proton cyclotron wave (PCW) activity, for time intervals when MAVEN was in the solar wind with no magnetic connection to the bow shock. It is shown that the nonlinear cascade of energy at the MHD scales is slightly amplified when PCWs are present in the plasma, \ADD{around the Martian perihelion}. In addition, the analysis of the normalized cross helicity and residual energy for the turbulent fluctuations shows the presence of Alfv\'enic and non-Alfv\'enic fluctuations in a magnetic dominant regime for the majority of the cases.
\end{abstract}

\section{Introduction}

Turbulence is a unique phenomenon present in several space environments, like the solar corona \citep{H1996,Dm2002}, planetary environments \citep{S2020} or the solar wind \citep{B2005,M2011}. In particular, solar wind turbulence is partially characterized by an inertial range, where energy is transferred without dissipation through different spatial and temporal scales \citep[e.g.,][]{F1995}. Typically, in the largest magnetohydrodynamic (MHD) scales, the solar wind magnetic spectrum presents a $-5/3$ slope \citep{K1941b,K1941a,M1982,L1998,Ch2016}, which is generally compatible with a constant energy cascade rate as a function of such scales \citep{SV2007,M2008,Co2014,Co2015,H2017a}. A constant energy cascade rate reflects a well accepted idea that large (MHD) scale turbulence serves as a reservoir of energy that cascades down to the smallest scales, where it can be dissipated by kinetic effects \citep[e.g.,][]{L1998,Sa2009,A2009,A2014b}. 

Assuming spatial homogeneity and full isotropy, an exact relation for fully developed incompressible MHD turbulence can be derived \citep{P1998b,P1998a}. Among its potential applications \citep[e.g.,][]{WEY2007,M1999,Mc2008,R1993,G1997,A2019}, the exact relation provides a precise computation of the amount of energy per unit time and volume (or heating rate) as a function of the velocity and magnetic correlation functions. The MHD exact relation and its connection with the nonlinear energy cascade rate has been numerically validated for both incompressible and compressible MHD turbulence \citep{A2018b}, and has been generalized to include sub-ion scale effects \citep{A2018,A2019b,H2017b,H2018,F2019,B2020}. Estimations of the energy cascade rate in the inertial range of solar wind turbulence have been previously computed at 1 Astronomical Unit (AU) \citep[see,][]{M2008,Co2014,Co2015,B2016c,H2017a} and more recently at $\sim$ 0.2 AU \citep{Ba2020,C2020}. In particular, \citet{H2017a} have investigated in detail the role of the compressible fluctuations \citep{B2013,A2017b} in modifying the energy cascade rate with respect to the prediction of the incompressible MHD model, based in situ data from the THEMIS/ARTEMIS spacecraft in the fast and slow solar wind.

The induced magnetosphere of Mars is formed as a result of the interaction between the solar wind and the planet's atmosphere, including its exosphere, ionosphere and the crustal magnetic fields \citep{A1998,A1999}. This interaction starts upstream of the Martian bow shock, due to the lack of an intrinsic global planetary magnetic field and the presence of an extended hydrogen exosphere \citep[e.g.,][]{Cha2015}. The response of this atmospheric obstacle is significantly modified by time-dependent physical processes \citep[e.g.,][]{E2010,J2015b,R2018a}, as a result of temporal variability of the planetary and solar wind properties over different timescales \citep[e.g.,][]{E2009,M2012,Ma2014,F2015,R2018b,R2019}. 

The seasonal variability of the Martian hydrogen exosphere has been identified by several spacecraft \citep{B2015,Ch2014,Cl2014,Cl2017,Ha2017a}. In particular, such variability exhibits higher column densities mostly observed around the Martian perihelion and lower column densities around aphelion \citep{Ha2017b}. The  Martian exosphere is subject to several ionizing mechanisms giving rise to newborn planetary protons, allowing one to indirectly observe such seasonal dependence with plasma instruments \citep[e.g.,][]{Y2015,R2017}. For instance, \citet{Y2015} reported a strong correlation between the detection rate of pickup ions originating from ionized exospheric hydrogen and the Martian heliocentric distance, based on Mars Express Ion Mass Analyzer observations. Higher pickup ions detection rates were observed when Mars is near perihelion. Moreover, when available, the measurement of the resulting proton velocity distribution function at these altitudes is composed of a core of solar wind particles and a non thermal proton population due to the presence of newborn planetary ions (seen in the solar wind reference frame). Such particle velocity distribution function is highly unstable and can give rise to several ultra-low frequency plasma waves \citep[][]{W1972,W1974,B1991,G1991,M1993,C2012}.

Despite their capability to excite different plasma waves, the relative velocity between the newborn planetary proton reference frame (very close to the planetary and spacecraft rest frames) and the solar wind is also responsible for a Doppler shift that defines the observed wave frequency near the local proton cyclotron frequency in the spacecraft reference frame \citep[e.g.,][]{R1990,B2002,M2004,R2013,Ro2016,Ru2015,Ru2016,L2020}. These waves are therefore called proton cyclotron waves (PCWs). Variability in the PCWs occurrence rate has been observed based on Mars Global Surveyor magnetic field data \citep{R2013,Be2013} and more recently with MAVEN Magnetometer (MAG) observations \citep{Ro2016,J2015,Con2015}. In particular, \citet{Ro2016} have analyzed MAG observations between October 2014 and March 2016. The authors reported that the PCWs occurrence rate upstream of the Martian bow shock varies with time and takes higher values near the Martian perihelion. Such long term trend was associated with higher hydrogen exospheric densities around that orbital position (derived from numerical simulations) and was also in agreement with the long term trend observed in the irradiances in the 121-122 nm range by MAVEN extreme ultra-violet monitor (EUVM)  measurements \citep{E2015}, which provide a proxy to study the temporal variability of the photoionization frequency of the neutral H exosphere. 

\citet{Ru2017} have characterized magnetic energy spectra in the Mars plasma environment using the MAVEN MAG observations, in the frequency range 0.005 Hz to 16 Hz. By computing the spectral indices for the magnetic energy, the authors showed a wide range of values in the upstream solar wind and the magnetosheath plasma. Also, they observed a seasonal variability of the spectral indices, indicative of a clear connection with the seasonal variability of the PCWs. Nevertheless, to the best of our knowledge, no estimation of the energy cascade rate has been reported yet in the Martian plasma environment. In the present Letter, we aim to extend the current state of knowledge of the solar wind turbulence upstream the Martian shock by computing for the first time the energy transfer rate using an exact relation for fully development turbulence. Using both magnetic field and plasma moments observations at $\sim 1.38-1.67$ AU, we investigate how turbulence is affected not only by the heliocentric distance, but also by the presence of PCWs. The study is structured as follows: in Section \ref{sec:model}, we present the theoretical incompressible MHD model and a brief description of the exact relation. In Section \ref{subsec:data} and \ref{subsec:criteria} we briefly describe the capabilities of the MAVEN instruments and the conditions that each turbulent event must fulfil, respectively. In Sections \ref{subsec:psd}, \ref{subsec:cascade} and \ref{subsec:alf} we present the main results of our analysis. Finally, the discussion and conclusions are developed in Section \ref{sec:discussion}.

\section{Incompressible MHD Turbulence}\label{sec:model}

The three-dimensional (3D) incompressible MHD equations are the momentum equation for the velocity field {\bf u} (in which the Lorentz force is included), the induction equation for the magnetic field {\bf B}, and the solenoid condition for both fields. These equations can be written as,
%
%\begin{strip}
\begin{align}\label{1} 
	&\frac{\partial \textbf{u}}{\partial t} = -\uh\cdot\boldsymbol\nabla\uh  + \ua\cdot\boldsymbol\nabla\ua - \frac{1}{\rho_0}\boldsymbol\nabla(P+P_M) + \textbf{f}_k  + \textbf{d}_k , \\ 	\label{2} 
    &\frac{\partial \ua}{\partial t} = - \uh\cdot\boldsymbol\nabla\ua + \ua\cdot\boldsymbol\nabla\uh + \textbf{f}_m + \textbf{d}_m , \\  \label{3} 
    &\boldsymbol\nabla\cdot\uh = 0, \\ \label{4} 
    &\boldsymbol\nabla\cdot\ua= 0
 \end{align}
%\end{strip}
%
where we have defined the incompressible Alfv\'en velocity $\ua\equiv\textbf{B}/\sqrt{4\pi\rho_0}$ (where $\rho_0$ the mean mass density) and $P_M\equiv\rho_0 u_\text{A}^2/2$ is the magnetic pressure. Then, both field variables, $\uh$ and $\ua$, are expressed in speed units. Finally, \textbf{f}$_{k,m}$ are respectively a mechanical and the curl of the electromotive large-scale forcings, and $\textbf{d}_{k,m}$ are respectively the small-scale kinetic and magnetic dissipation terms \citep{A2016b,B2018}. 

Using Eq.~\eqref{1}-\eqref{4} and following the usual assumptions for fully developed homogeneous turbulence (i.e., infinite kinetic and magnetic Reynolds numbers and a steady state with a balance between forcing and dissipation \citep[see, e.g.][]{A2017b}, an exact relation for incompressible MHD turbulence can be obtained as \citep{P1998b,P1998a},
\begin{align}\label{exactlaw0}
	-4\varepsilon&= \rho_0\boldsymbol\nabla_\ell\cdot\langle (\delta\uh\cdot\delta\uh+\delta\ua\cdot\delta\ua)\delta\uh - (\delta\uh\cdot\delta\ua+\delta\ua\cdot\delta\uh)\delta\ua\rangle,
\end{align}
where $\varepsilon$ is the total energy cascade rate per unit volume. Fields are evaluated at position $\textbf{x}$ or $\textbf{x}'=\textbf{x}+\boldsymbol\ell$; in the latter case a prime is added to the field. The angular bracket $\langle\cdot\rangle$ denotes an ensemble average \citep{Ba1953}, which is taken here as time average assuming ergodicity. Finally, we have introduced the usual increments definition, i.e., $\delta\alpha\equiv\alpha'-\alpha$. Here we are interested in estimating $\varepsilon$ from Eq.~\eqref{exactlaw}, which is fully defined by velocity and magnetic field increments (or fluctuations) that we can estimate from MAVEN observations.

\section{Analysis and Results}\label{sec:results}

\subsection{MAVEN observations}\label{subsec:data}

The MAVEN spacecraft was launched in November 2013 and arrived at Mars in September 2014. The orbit had a nominal periapsis altitude of 150 km, an apoapsis altitude of 6220 km, and an orbital period of about 4.5 hours \citep{J2015}. The selected apoapsis altitude, orbital period and inclination (75$^\circ$) allow orbital precession in both local time and latitude of the spacecraft periapsis. In addition, the extent of the MAVEN apoapsis allows sampling of solar wind properties upstream of its bow shock. The MAVEN Magnetometer (MAG) provides vector magnetic field measurements with a 32 Hz maximum sampling frequency and absolute vector accuracy of 0.05$\%$ \citep{Con2015}. MAVEN’s Solar Wind Ion Analyzer (SWIA) is an energy and angular ion spectrometer covering an energy range between 25 eV/q and 25 keV/q with a field of view of 360$^\circ\times$90$^\circ$ \citep{Ha2015b}. In this study, we have analyzed MAVEN MAG and SWIA data sets as follows. Magnetic field observations with 32 Hz cadence are analyzed to discriminate events in the pristine solar wind with PCWs and without wave activity. To estimate the energy cascade rate at MHD scales (i.e., frequencies below $\sim 0.1$ Hz) we averaged the magnetic field data over 4 s to match SWIA onboard moments cadence \citep{Ha2015b}. In particular, in this study we have used SWIA solar wind density and velocity onboard computed moments. Such moments assume a plasma made of protons, a very good approximation upstream from the Martian bow shock \citep{Ha2017a}.

As discussed in the Introduction, \citet{Ro2016} have found that the PCWs occurrence rate increases (up to $\sim$ 50$\%$) when Mars is close to the perihelion (1.38 AU) on December 12 2014 and remains relatively low and constant ($\sim$ 25$\%$) after reaching the Martian Northern Spring Equinox-Southern Autumn Equinox (NSE-SAE). Also, the authors concluded that the increment in the PCWs occurrence rate cannot be the result of biases associated with MAVEN’s spatial coverage of the upstream region or of the differences in the spatial distribution of the crustal magnetic fields. Therefore, to investigate how PCWs activity may affect the nonlinear transfer of energy, we consider two data sets. Set A contains observations from December 1 2014 until January 31 2015; and set B from January 1, \ADD{2016} until February 29, 2016. Set A includes MAVEN observations around perihelion and a local maximum of PCWs activity, while set B includes the Martian Northern Summer Solstice-Southern Winter Solstice (NSS-SWS) that took place on January 3 2016 (and corresponds to a local minimum of waves activity).

\subsection{Selection criteria}\label{subsec:criteria}

For sets A and B ($\sim$ 330 orbits per set), during time periods when MAVEN was traveling in the solar wind with no connection to the shock \citep{Gru2018}, we looked for intervals in which the number density fluctuation level was lower than $20\%$ (to be as close as possible to the incompressibilty condition). Moreover, in order to have a reliable estimate of the energy cascade rate $\varepsilon$ (both its sign and its absolute value \citep{Ha2017a}) we only consider the events in which the $\theta_{uB}$ (the angle between the magnetic and velocity field) was relatively stationary \citep{A2019b}. The time interval of each MAVEN orbit that fulfill these conditions were divided into a series of sample events with a duration of 30 minutes. This duration ensures having at least one correlation time of the turbulent fluctuations \citep{H2017a,Ma2018}. Finally, for set A (set B) we considered only cases when PCWs activity was present (absent). By doing this, we can assess the effects that the PCWs may have on the solar wind turbulence. This selection resulted in 184 and 208 events for sets A and B, respectively. 

\begin{figure*}
\begin{center}
\includegraphics[width=1\textwidth]{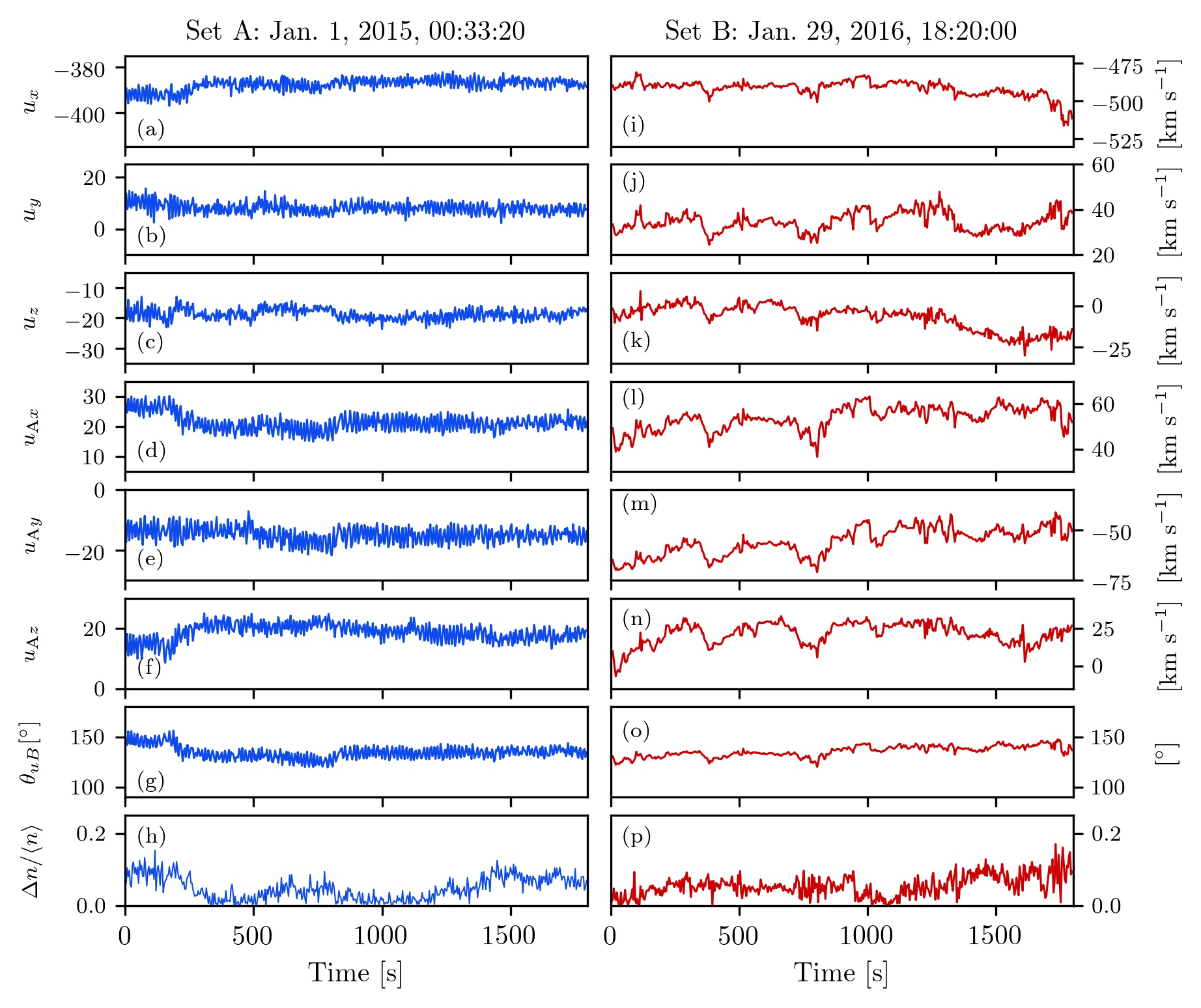}
\end{center}
\caption{Time  series for two examples from sets A and B. In particular, the proton and Alfv\'en velocity field components (in MSO coordinate system), the angle between magnetic and velocity fields and the density fluctuation level, respectively.}
\label{fig0}
\end{figure*}

Figure \ref{fig0} shows two examples of the typical events analyzed in the present Letter (panels (a)-(h) show an example from set A, and panels (i)-(p) from set B). Figure \ref{fig0} (a)-(f) show the time series for the proton and Alfv\'en velocity field components in Mars-centered Solar Orbital (MSO) coordinate system (where the {\bf x}-axis points from Mars to the Sun, {\bf z}-axis is perpendicular to Mars' orbital plane and is positive toward the ecliptic north; the {\bf y}-axis completes the right-handed system). Figure \ref{fig0} (g)-(h) show the angle between the magnetic and velocity field $\theta_{uB}$ and the density fluctuation level (i.e., $\Delta n/\langle n\rangle$), respectively. As can be seen, although both examples show approximately the same level of density fluctuations and the same $\theta_{uB}$ angle, there is a sharp contrast with and without clear wave activity present in the left and right panel, respectively.
%Finally, the Supporting Information shows that both sets A and B have similar distributions for the density, velocity and magnetic fluctuation values.

\begin{figure*}
\begin{center}
\includegraphics[width=0.95\textwidth]{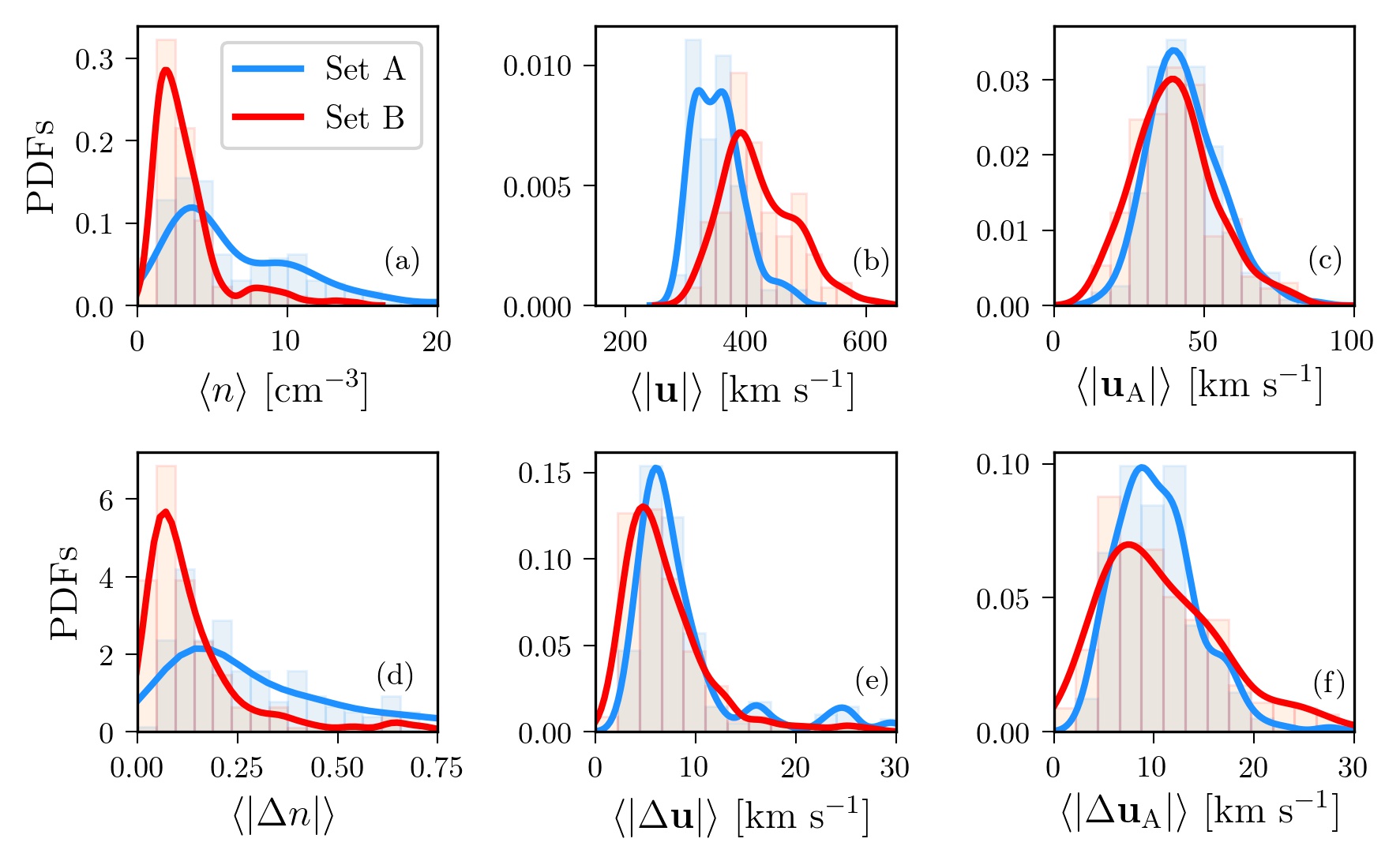}
\end{center}
\caption{Probability distribution functions for (a)-(c) the number density, velocity and Alfv\'en velocity fields absolute values, respectively. Panels (d)-(f) show the PDFs for the fluctuation values.}
\label{fig1}
\end{figure*}

Figure \ref{fig1} shows the probability distribution functions (PDFs)  for all the analyzed events in both sets for (a)-(c) the number density, velocity, and Alfv\'en velocity field absolute values and for (d)-(f) its fluctuation amplitudes, respectively. Both sets have similar distributions for the fluctuation values.

\subsection{PSD of the magnetic field fluctuations}\label{subsec:psd}

To determine if a given time interval presents PCWs activity or not, we used a criterion similar to the one in \citet{Ro2016}. An event is considered to present PCWs activity when the magnetic energy power spectral density (PSD) displays an increase in a frequency interval centered around the local proton cyclotron frequency $f_{ci}$ when compared to two contiguous windows of width 0.2 $f_{ci}$. More precisely,
\begin{equation}
{\it max}\{\text{PSD}[B(f)]|^{1.2f_{ci}}_{0.8f_{ci}}\} > {\it max}\{\text{PSD}[B(f)]|^{1.4f_{ci}}_{1.2f_{ci}}\}, {\it max}\{\text{PSD}[B(f)]|^{0.8f_{ci}}_{0.6f_{ci}}\}
\end{equation}
where {\it max} corresponds to the maximum value in the PSD in the corresponding window. 

\begin{figure*}
\begin{center}
\includegraphics[width=0.95\textwidth]{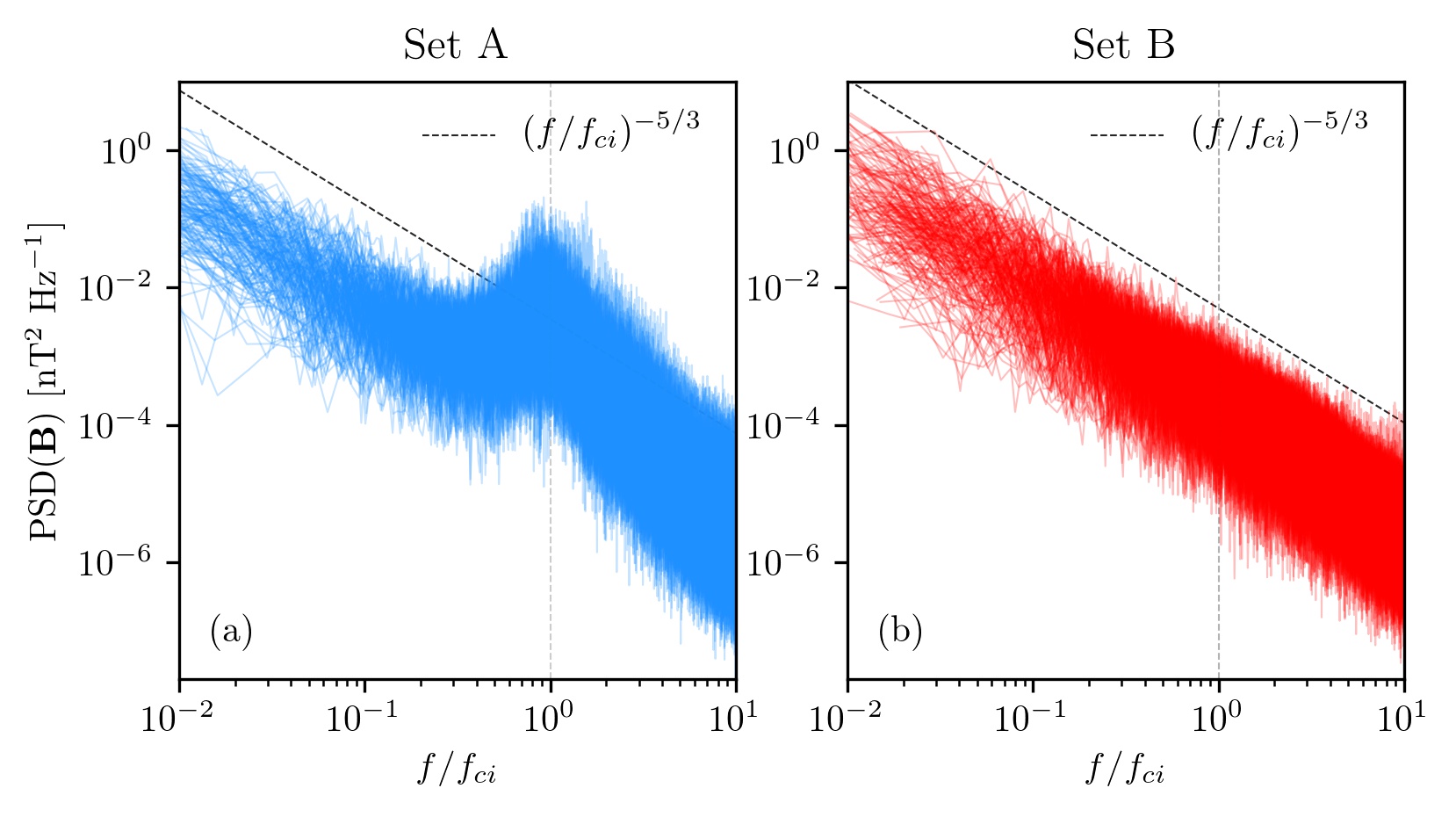}
\end{center}
\caption{Magnetic power spectra density for both sets A and B, respectively.}
\label{fig2}
\end{figure*}

Figure \ref{fig2} (a) and (b) show the PSD for all the events in sets A and B, respectively. For reference, we plot a straight line with Kolmogorov-like slope (i.e., -5/3) in both cases. As we expected, all events near the Martian perihelion (i.e., set A) show a clear peak in their PSD near the proton cyclotron frequency $f_{ci}$. Moreover, all the cases analyzed in the present Letter show a Kolmogorov-like slope in the MHD scales \citep[see,][]{Ru2017}. Figure \ref{fig3} (a) and (b) show the MAVEN location (where R$_\text{MSO}=\sqrt{y_\text{MSO}^2+z_\text{MSO}^2}$) for each event for sets A and B, respectively. The gray dashed line corresponds to the best fit of the bow shock reported by \citet{Gru2018}. In that study, the authors presented a model of the Martian bow shock as a three dimensional surface making use of 1000 crossings observed by MAVEN. As shown in Figure \ref{fig3} insets, sets A and B present some differences in the cylindrical MSO spatial distributions. In particular, the statistical distribution of observations corresponding to set B are slightly closer to the Martian bow shock fit.

\subsection{Energy cascade rates}\label{subsec:cascade}

To compute the right hand side of Eq.~\eqref{exactlaw0}, we constructed temporal correlation functions of the different turbulent fields at different time lags $\tau$ in the interval [4,1800] s, which allows covering the MHD inertial range \citep{Ru2017,H2017a}. \ADD{More precisely, assuming isotropic turbulence and the Taylor hypothesis (i.e., $\ell\equiv\tau V$, where $V$ is the mean plasma flow speed and $\ell=|\boldsymbol\ell|$ is the longitudinal distance), Eq.~\eqref{exactlaw0} can be expressed as a function of time lags $\tau$. Since we are a dealing with single spacecraft measurements, we assume that the isotropic energy cascade rate is representative of the real cascade rate. In particular, while Eq.~\eqref{exactlaw0} includes increments in all the spatial directions, here we only include the increments in the longitudinal direction $\ell$. Therefore, for each event in both sets, the energy cascade rate is computed as,
\begin{align}\label{exactlaw}
	\varepsilon&= \rho_0\langle [(\delta\uh\cdot\delta\uh+\delta\ua\cdot\delta\ua)\delta{u}_\ell - (\delta\uh\cdot\delta\ua+\delta\ua\cdot\delta\uh)\delta{u}_{\text{A}\ell}]/(-4/3 \ell)\rangle.
\end{align}
}

Figure \ref{fig4} (a) and (b) show the absolute value of the energy cascade rate as a function of the time lag ($\tau$) for both sets. Figure \ref{fig4} (c) shows the histogram for the (log) mean values $\log\langle|\varepsilon|\rangle_\mathrm{MHD}$ in the MHD scales ($\tau=5\times10^2-1.5\times10^3$ s). It is worth emphasizing that if $\varepsilon$ is changing significantly in amplitude and/or sign, then the resulting mean values would not be reliable \citep[see, e.g.,][]{H2017b,A2019b}. Therefore, as we mentioned before, we kept only the intervals for which the cascade rate shows a constant (negative or positive) sign for all the time lags in the MHD range. By doing so, the mean value of $\varepsilon$ for each event is robust and so is its absolute value \citep{Co2015,H2017b}. The only limitation of analyzing the non-signed $\varepsilon$ is related to the direct vs.~inverse nature of the energy cascade rate. This is because the convergence of the sign of $\varepsilon$ is more stringent than its absolute value \citep[see,][]{Co2015,H2017b}, thus demanding a much larger statistical sample than the one considered in the present work. For both data sets A and B, the cascade rate values are lower than the averaged value observed at 1 AU, $\varepsilon\sim 10^{-16}-10^{-17}$ J m$^{-3}$ s$^{-1}$ \citep{H2017b}. Also, it is worth mentioning that the energy cascade rate increases slightly when PCWs are present in the solar wind, based on our statistical analysis.

\subsection{Alfv\'enic fluctuations}\label{subsec:alf}

The cross helicity $H_c = \langle\uh\cdot{\ua}\rangle$ and the total energy $E_T\equiv(\langle |\uh|^2 \rangle+\langle |\ua|^2\rangle)/2$ (where $\uh$ and $\ua$ are the proton and Alf\'en velocities fluctuations) are the two rugged invariant of the ideal incompressible MHD model (see Eqs.~\ref{1}-\ref{4}). The dimensionless measure of the normalized cross-helicity corresponds to $\sigma_c\equiv H_C/E_T$, with $-1\leq\sigma_{c}\leq1$. Usually, fluctuations with $|\sigma_c|\sim1$ are described as being Alfvénic. Another related measurement to quantify the relative energy present in the kinetic and magnetic fluctuations is the normalized residual energy $\sigma_r\equiv(\langle |{\bf u}|^2 \rangle-\langle |{\bf u}_\text{A}|^2\rangle)/E_T$. This parameter also ranges between -1 and 1.

Figure \ref{fig5} shows the scatter plot of $\sigma_r$ as a function of $\sigma_c$, for both sets A and B, respectively. The colorbar corresponds to the mean value of the energy cascade rate in the MHD scales $\langle|\varepsilon|\rangle_\text{MHD}$. The statistical results show a wide variety of possible values of $\sigma_r$ and $\sigma_c$, independently of the presence of PCWs. However, for set B, the events gather around $|\sigma_c|\sim0.75$ and $\sigma_r\sim-0.4$.

\section{Discussions and Conclusions}\label{sec:discussion}

\begin{figure*}
\begin{center}
\includegraphics[width=0.75\textwidth]{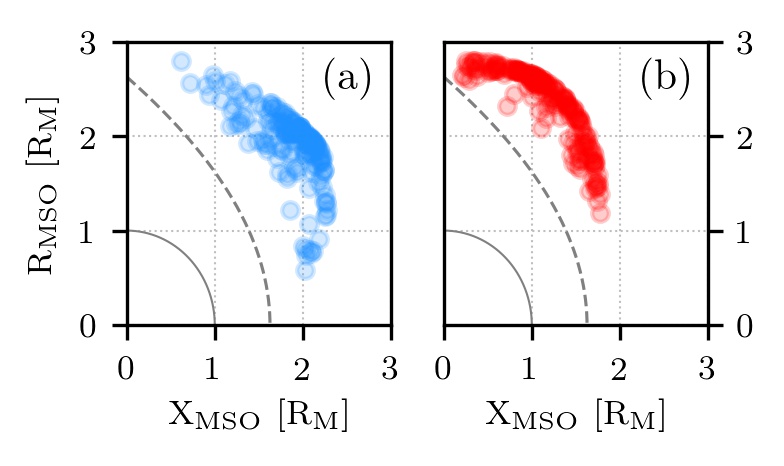}
\end{center}
\caption{MAVEN location of each event in MSO reference frame for both sets. The dash gray line is the bow shock best fit from \citet{Gru2018}.}
\label{fig3}
\end{figure*}

\begin{figure}
\begin{center}
\includegraphics[width=0.95\textwidth]{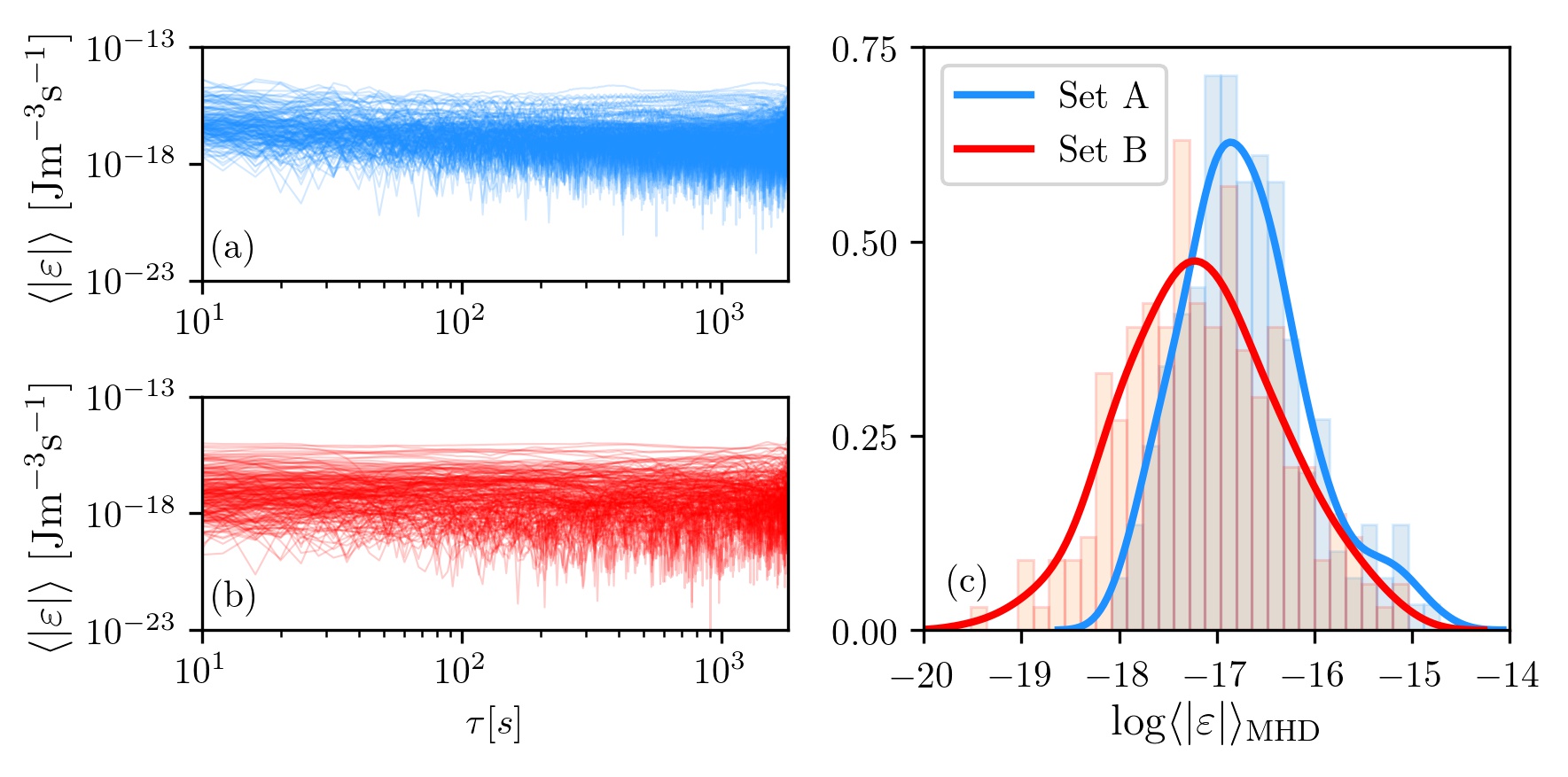}
\end{center}
\caption{Energy cascade rate (absolute value) as a function of the time lag for sets (a) A and (b) B, respectively. (c) Histogram of $\log\langle|\varepsilon|\rangle_\text{MHD}$ for both sets.}
\label{fig4}
\end{figure}

\begin{figure}
\begin{center}
\includegraphics[width=0.95\textwidth]{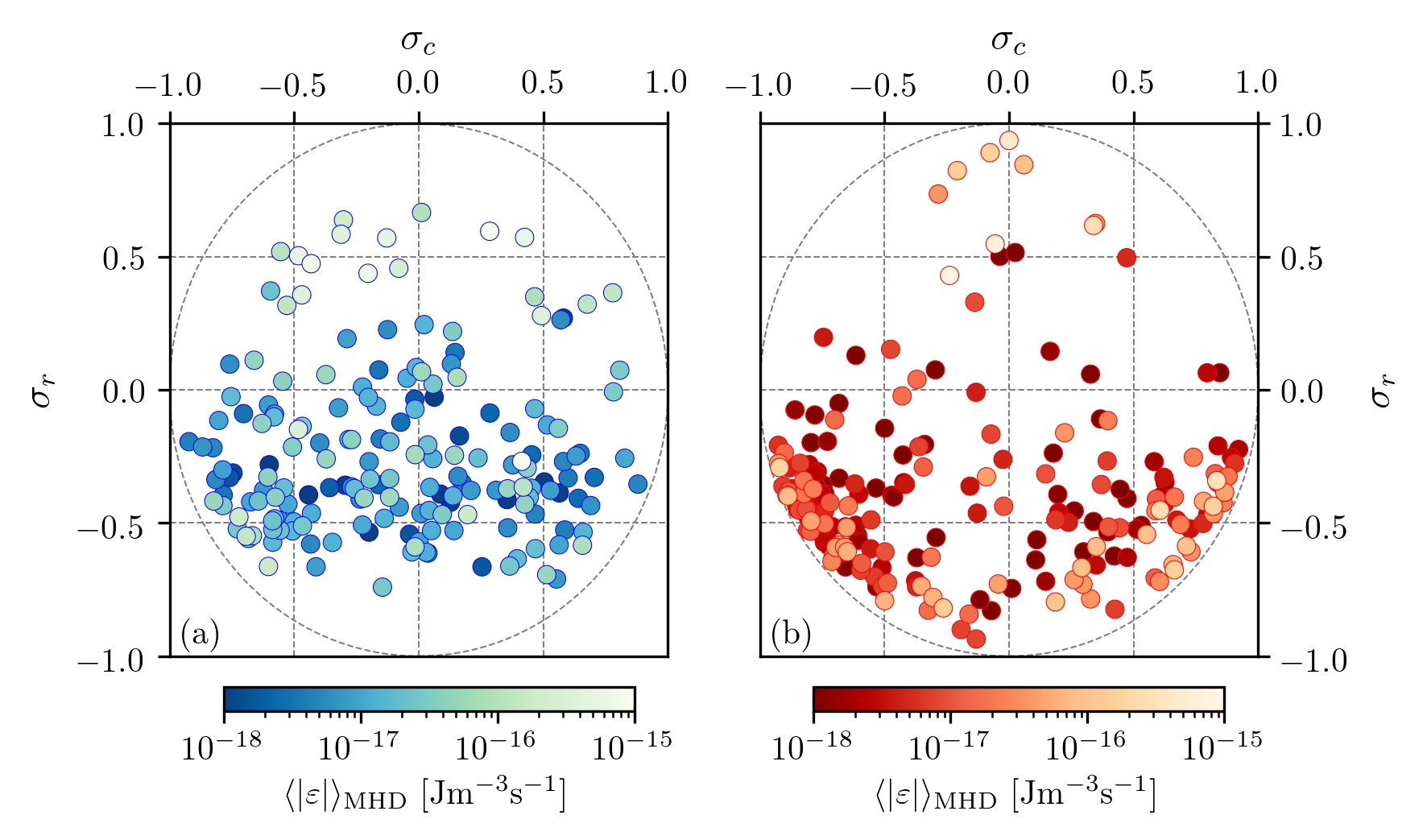}
\end{center}
\caption{Scatter plot of $\sigma_r$ as a function of $\sigma_r$ for both sets A and B, respectively. Color bars correspond to the mean cascade rate in the MHD scales.}
\label{fig5}
\end{figure}

In the present work, we analyzed two data sets by considering separately the cases with (set A) and without PCWs (set B). In agreement with previous studies, our findings are consistent with the seasonal variability of PCWs \citep{R2013,Be2013,Ro2016}. As shown in Figure \ref{fig3}, the spatial distributions of the cylindrical MSO coordinates associated with the events in sets A and B display some differences. However, as reported in \citet{Ro2016}, the PCWs occurrence rate temporal variability upstream from the bow shock cannot be associated with biases in the spatial coverage of MAVEN or with changes in the background magnetic fields. Moreover, in this work we restricted our analysis to time intervals with no connection to the Martian bow shock to avoid effects associated with backstreaming particles \citep{Maz2018}. 
Our statistical results show slopes compatible with a Kolmogorov scaling in the largest MHD scales in both sets. \citet{Ru2017} determined spectra of magnetic field fluctuations in order to characterize turbulence in the Mars plasma environment. Using 512 s sliding windows, the authors found that magnetic spectrum slopes present different values. In particular, they found that the slope is typically $\sim - 1.2$ at the solar wind (in the MHD scales), which differs from the Kolmogorov spectrum. This discrepancy between the computed slopes could be due to several factors: i) we are including only the cases where the cascade rate and the angle $\theta_{uB}$ are approximately constant; ii) the sliding window size used in \citet{Ru2017} may not include enough correlation times to yield reliable PSD slopes; and iii) we are separating between PCWs and no waves events, while \citet{Ru2017} included all the available data. It is worth mentioning that \citet{G2010} showed that the magnetic field fluctuations have a Kolmogorov scaling using magnetic field values derived from electron cyclotron echoes from Mars Express observations. Also, the $f^{-5/3}$ spectrum for the magnetic energy is theoretically compatible with our constant energy cascade rate assumption \citep{A2016b,A2016c}. 

We found that the energy cascade rate at Mars ($\sim$ 1.38 $-$ 1.67 AU) decreases compared with previous results at 1 AU and smaller distances from the Sun \citep[see,][]{M2011,B2005,H2017a,Ba2020}. In particular, the statistical results for the data set B (no presence of PCWs activity) show a decrease of $|\varepsilon|$ of at least 1 order of magnitude with respect to the value at 1 AU (i.e., $10^{-16}-10^{-17}$ J m$^{-3}$ s$^{-1}$) \citep{H2017b}. Recent observations have shown that the energy cascade rate becomes larger closer to the Sun \citep{Ba2020}. Probably, these large values of the energy cascade rate are due to strong forcing process occurring closer to the Sun's corona. However, a precise study comparing the energy cascade rate at other different heliocentric distances can be found elsewhere \citep[e.g.,][]{B2005}. 

In addition, we observe a slight increase in the transfer of energy when waves activity is present in the plasma (set A). These results suggest that PCWs at the sub ion scales may affect the turbulence properties at the MHD scales. In other words, while Eq.~\eqref{exactlaw} is valid {\it only} in the MHD inertial range, our results suggest that the instabilities and consequent nonlinear waves at frequencies $\sim f_{ci}$ may affect the largest MHD scales \citep{O2013,H2017b}. It is worth mentioning that, although several theoretical papers have shown that newborn planetary ions are capable of providing the free energy for the presence of PCWs \citep[e.g.,][]{B1991}, the PCWs observed upstream from the Martian bow shock are nonlinear and likely not saturated \citep{C2012}. Thus, while the observed increase in $|\varepsilon|$ in correlation with PCWs activity has not been reported before, an analysis of the local velocity distribution functions is still needed to better characterize the growing stage of the observed PCWs and its connection with these results \citep[e.g.,][]{M2000,M2003,R2018c}. Moreover, a comparative analysis of these energy cascade rate estimates with the ones upstream from the bow shocks of Venus and active comets would allow us to better determine the significance of such correlation. Indeed, waves with frequencies close to the local proton and water group ion cyclotron frequency have been reported upstream from Venus and several active comets, respectively, and therefore might also affect the pristine solar wind energy cascade rates \citep[see,][]{M1997,D2015}.

While both sets show similar values in the parameter space of number density, velocity and Alfv\'en velocity fields values, our results show a wide variability in the possible values of $\sigma_c$ and $\sigma_r$. In particular, the events in set B correspond to Alfv\'enic and magnetic dominant fluctuations ($|\sigma_c|\sim0.75$ and $\sigma_r\sim-0.4$). Interestingly, these events correspond to the higher values of the cascade rate in the set B. Moreover, for both sets the events have mainly negative $\sigma_r$ values with a majority gathering around $ \sigma_r\sim-0.25$ and $\sigma_r\sim-0.4$, respectively. This majority of events in the magnetic dominant regime is compatible with previous results between 1 and 8 AU \citep{Ro1990,B2007,Ma2008,Ha2017a}. In particular, \citet{Ha2017a} have investigated the spatial distributions of $\sigma_r$ and $\sigma_c$ using 30 minutes time intervals with a 45 s cadence. Separating observations into four subsets based on the $B_y$ sign and the time range (near perihelion or aphelion), the authors found that the temporal decrease in $\sigma_c$ appears to be equally present in all upstream regions sampled by MAVEN. Our results using 4 s or 45 s (not shown here) cadence exhibit a similar statistical trend. Therefore, the PCWs activity is not affecting significantly the mean value of the statistical distributions of $\sigma_r$ and $\sigma_c$. Slight differences with \citet{Ha2017a} are probably due to the considered selection criteria.

Finally, in this study we have not computed the compressible component of the energy cascade rate \citep{B2013,A2017b}. In particular, we expect to obtain a strong increase in the nonlinear cascade rate of energy in the Martian magnetosheath, where compressibility plays a major role, like in the Earth's magnetosheath \citep{H2017b,A2019b}. \ADD{In addition, we emphasize that we can not definitively state that the presence of PCWs is the ultimate responsible for the energy cascade rate increase. Indeed, another possible reason that could explain the observed increase in the cascade rate amplitude might be the heliocentric distance variation along Mars orbit. The presence of waves, as the energy cascade rate amplitude, are both dependent on the Martian heliocentric distance. Therefore, to decouple effects due to the presence of PCWs, the cascade rate amplitude and the heliocentric distance is necessary to consider additional magnetic field and plasma data sets. Furthermore, a possible seasonal variability of the incompressible and/or compressible energy cascade rate may be present in the Martian environment as well. These studies will be part of an ongoing work.}

\acknowledgments
N.A., L.H.Z. and F.S.~acknowledge financial support from CNRS/CONICET Laboratoire International Associ\'e (LIA) MAGNETO. We thank the entire MAVEN team and instrument leads for data access and support.  N.A. acknowledge financial support from the Agencia de Promoci\'on Cient\'ifica y Tecnol\'ogica (Argentina) through grants PICT 2018 1095. MAVEN data are publicly available through the Planetary Data System (\url{https://pds-ppi.igpp.ucla.edu/index.jsp}).

\bibliography{cites}{}
\bibliographystyle{aasjournal}

\newpage

\end{document}